\documentclass[a4paper]{article}

\usepackage{INTERSPEECH2022}
\usepackage{color}
\usepackage{threeparttable}
\usepackage{multirow}
\usepackage{booktabs}

\title{Multi-Channel Far-Field Speaker Verification with Large-Scale Ad-hoc Microphone Arrays}
\name{Chengdong Liang, Yijiang Chen, Jiadi Yao, Xiao-Lei Zhang}
\address{
CIAIC, School of Marine Science and Technology, Northwestern Polytechnical University, Xi'an, China}
\email{liangchengdong@mail.nwpu.edu.cn, xiaolei.zhang@nwpu.edu.cn}
\begin{document}

\maketitle
\begin{abstract}
Speaker verification based on ad-hoc microphone arrays has the potential of reducing the error significantly in adverse acoustic environments. However, existing approaches extract utterance-level speaker embeddings from each channel of an ad-hoc microphone array, which does not consider fully the spatial-temporal information across the devices. In this paper, we propose to aggregate the multichannel signals of the ad-hoc microphone array at the frame-level by exploring the cross-channel information deeply with two attention mechanisms. The first one is a self-attention method. It consists of a cross-frame self-attention layer and a cross-channel self-attention layer successively, both working at the frame level. The second one learns the cross-frame and cross-channel information via two graph attention layers. Experimental results demonstrate that the proposed methods reach the state-of-the-art performance. Moreover, the graph-attention method is better than the self-attention method in most cases.

\end{abstract}

\section{Introduction}

With the rapid popularization of smart terminal devices, such as vehicle-mounted devices, smart speakers at home, etc., far-field automatic speaker verification (ASV) is commonly required. However, due to the noise, reverberation and speech signal attenuation in far-field environments, the performance of single-channel ASV drops sharply. To make the smart devices robust in the noise and reverberant environments, many single-channel and  multi-channel speech enhancement based on a single device has been extensively studied \cite{taherian2020robust,jiang2014binaural,wang2018all,abd2020text}. For example, beamforming, which explores the spectral and spatial diversity of the target and interference signals via multiple microphones at a single device \cite{movsner2018dereverberation,jin2010speaker,taherian2019deep}, leads to substantial performance improvement of ASV.

Recently, a novel kind of front-ends, named ad-hoc microphone arrays, have received attention as an alternative solution to the far-field speech processing.
Unlike the fixed microphone array at a single device, an ad-hoc microphone array is composed of a group of microphone devices/nodes randomly distributed in space, where the number and arrangement of the nodes may be unknown in practice \cite{zhang2021deep}. It allows users to use their own devices to virtually form a microphone array system flexibly. It has the potential to reduce the occurrence probability of the far-field problem. It also captures more spatial and temporal information than the fixed microphone array at a single device.

Recently, several studies on ad-hoc microphone arrays have been conducted. In \cite{yang2020deep}, the authors proposed deep ad-hoc beamforming based on speaker extraction for speech separation. It uses a supervised channel selection framework and a deep learning based MVDR algorithm to extract the targeted speech from a mixed speech of multiple speakers. In \cite{wang2021continuous,wang2020neural}, the authors used the attention mechanism to obtain the relevant information between and within channels for multi-channel speech separation. In \cite{guan2021libri}, the authors released a large-scale ad-hoc data set and a baseline system for speech recognition. In \cite{chang2021end,chang2021multi}, the authors applied neural transformer architectures for multi-channel speech recognition systems, where the multichannel speech signals are aggregated by attention layers. In \cite{chen2021scaling}, the authors further conducted channel selection by a scaling sparsemax operator for speech recognition with large ad-hoc arrays.

{With regard to the ASV study with ad-hoc microphone arrays, \cite{liang2021attention,cai2021embedding} independently proposed to aggregate the multi-channel speech signals at the utterance-level. They add an aggregation layer above the pooling layer to fuse the speaker embeddings from different channels. The difference between them is mainly on how to aggregate the speaker embeddings. \cite{liang2021attention} used cross-channel self-attention mechanism, while \cite{cai2021embedding} adopted attentive pooling layer.} An apparent drawback of the utterance-level aggregation methods in \cite{liang2021attention, cai2021embedding} is that it misses the spatial-temporal information between the channels.

To address this problem, in this paper, we propose to aggregate the multichannel signals of ad-hoc microphone arrays at the frame-level by adding the multi-channel aggregation module before the pooling layer. We propose two multi-channel aggregation modules. The first one is the \textit{self-attention based aggregation} (SA-aggregation), in which cross-frame and cross-channel self-attentions are stacked to capture the temporal and spatial information respectively. Unlike \cite{wang2020neural,wang2021continuous}, our SA-aggregation does not adopt feed-forward networks (FFN), since that they empirically degrade performance. A weakness of SA-aggregation is that it builds the connections between acoustic features indirectly, where the attention weight is calculated as the dot product of a feature and a projection vector instead of the calculation between the features.  As we know, the direct connections between the features are important information for the multi-channel aggregation.

Inspired by the recent success of graph neural networks \cite{velivckovic2017graph,brody2021attentive}, the second multi-channel aggregation module is based on graph attention (GAT-aggregation). It consists of a stack of the cross-frame and cross-channel graph attention layers. It obtains a single representation from the correlation between all possible microphone node pairs. It overcomes the weakness of SA-aggregation.

We conducted an extensive experiment on a simulated corpus generated from Librispeech as well as the semi-real Libri-adhoc40 \cite{guan2021libri} corpus. Experimental results with ad-hoc microphone arrays of as many as $40$ nodes demonstrate that the proposed SA-aggregation and GAT-aggregation outperforms the utterance-level methods in both data. With the increase of the number of ad-hoc nodes, the advantage of the proposed frame-level methods over the utterance-level methods is enlarged. Moreover, GAT-aggregation achieves better performance than SA-aggregation. For example, the best equal error rate (EER) of GAT-aggregation is $6.86\%$, which is $6.81\%$ relatively lower than that of SA-aggregation. The proposed methods trained with the simulated data shows better generalization to real scenarios than the utterance-level methods.

\section{Proposed method}

\begin{figure}[t]
	
	\centering
	\includegraphics[width=0.6\linewidth]{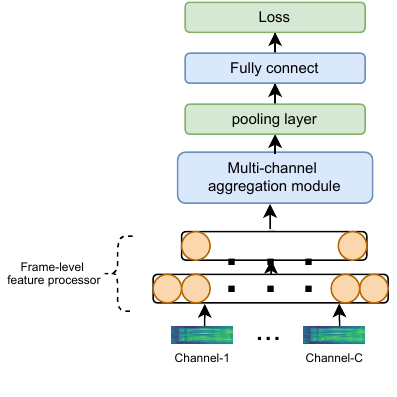}
	\caption{Architecture of the proposed multi-channel ASV system. The blue block is fine-tuned with ad-hoc data.}
	\label{fig:multi-channel}
\end{figure}

Figure~\ref{fig:multi-channel} shows the architecture of the proposed multi-channel frame-level speaker verification model with ad-hoc microphone arrays. It consists of two elements: a frame-level feature processor and a multi-channel aggregation module. 

\textbf{Note that}, although some mathematical symbols in Sections \ref{subsec:sa} and \ref{subsec:ga} are used duplicately across the subsections, they have different definitions in different subsections, since that the two subsections present the SA-aggregation and GAT-aggregation independently. The reason why we used the same symbols is to make readers easily understand the two algorithms comparatively which share a similar procedure.

{We denote $T$ as the number of frames, $C$ as the number of input channels, and $D$ as the feature dimension of each channel. The input of the multi-channel aggregation module is denoted as $\mathbf{X} = \left[\mathbf{X}_{0}, \cdots ,\mathbf{X}_{C-1}\right]$ as , where $\mathbf{X}_c \in  \mathbb{R}^{T \times D}$ is the input feature matrix of the $c$-th channel. Let $M$ denote the number of the attention heads.}

\subsection{Aggregation module based on self-attention}\label{subsec:sa}
As shown in Figure~\ref{fig:SA}, the aggregation module based on self-attention consists of a cross-frame self-attention layer and a cross-channel self-attention layer. The entire network can make use of information across channels and frames. The detailed calculation process is as follows.
\subsubsection{Cross-frame self-attention layer}
  For each attention head, the input features are transformed into query ($Q$), key ($K$) and value ($V$) subspaces of dimension $E$ as follows:
\begin{equation}
\mathbf{Q}_{c}^{m} =\mathbf{X}_{c}\mathbf{W}_{Q}^{m}, \mathbf{K}_{c}^{m} =\mathbf{X}_{c}\mathbf{W}_{K}^{m}, \mathbf{V}_{c}^{m} =\mathbf{X}_{c}\mathbf{W}_{V}^{m}
\label{eq:1}
\end{equation}
where $d_k = E / K$, the matrices $\mathbf{Q}_{c}^{m}$, $\mathbf{K}_{c}^{m}$, $\mathbf{V}_{c}^{m}$ denote the query, key, and value embeddings respectively, all of which are in $\mathbb{R}^{T \times d_k}$. For the $m$-th attention head at channel $c$, $\mathbf{W}_{*}^{m} \in \mathbb{R}^{D \times d_k}$ are the trainable parameters where $* \in \{K,Q,V\}$. The similarity matrix is computed as the product of the query and key matrices. 
The output of the $m$-th attention head is then computed by:
\begin{equation}
\mathbf{H}_{c}^{m}=\operatorname{softmax}\left(\frac{\mathbf{Q}_{c}^{m} \cdot \left(\mathbf{K}_{c}^{m}\right)^{\top}}{\sqrt{d_k}} \right) \mathbf{V}_{c}^{m}
\label{eq:2}
\end{equation}
where $\mathbf{H}_{c}^{m} \in \mathbb{R}^{T \times d_k}$. Finally, the output of all attention heads are concatenated across the subspaces by:
\begin{equation}
\mathbf{Y}_c=\operatorname{concat} \left[\mathbf{H}_{c}^{1}, \mathbf{H}_{c}^{2}, \ldots, \mathbf{H}_{c}^{M}\right] \mathbf{W}_{O}
\end{equation}
where $\mathbf{Y}_{c} \in \mathbb{R}^{T \times D}$ and $\mathbf{W}_{O} \in \mathbb{R}^{E \times D}$ is a weight matrix of the linear projection layer.

\subsubsection{Cross-channel self-attention layer}
\begin{figure}[t]
	
	\centering
	\includegraphics[width=0.6\linewidth]{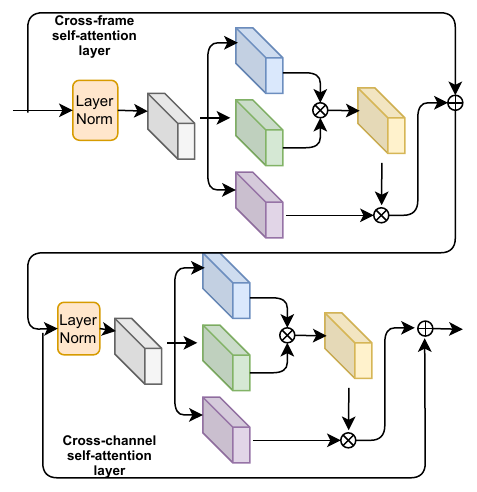}
	\caption{Architecture of the multi-channel aggregation based on self-attention.}
	\label{fig:SA}
\end{figure}

Then, we transform $\mathbf{Y} = \left[\mathbf{Y}_{0}, \cdots ,\mathbf{Y}_{C-1}\right]$ to $\mathbf{\hat X} = \left[\mathbf{\hat X}_{0}, \cdots ,\mathbf{\hat X}_{T-1}\right]$ as the input of the cross-channel self-attention, where $\mathbf{\hat X}_{t} \in \mathbb{R}^{C \times D}$. Similar to the cross-frame self-attention layer, the cross-channel self-attention can be denoted as follows:
\begin{equation}
\begin{aligned}
\mathbf{\hat H}_{t}^{m} &=\operatorname{softmax}\left(\frac{\mathbf{Q}_{t}^{m} \cdot \left(\mathbf{K}_{t}^{m}\right)^{\top}}{\sqrt{d_k}}\right) \mathbf{V}_{t}^{m} \\
\mathbf{\hat Y}_{t} &=\operatorname{concat} \left[\mathbf{\hat H}_{t}^{1}, \mathbf{\hat H}_{t}^{2}, \ldots, \mathbf{\hat H}_{t}^{M}\right] \mathbf{\hat W}_{O}
\end{aligned}
\end{equation}
where
\begin{equation}
\mathbf{Q}_{t}^{m} =\mathbf{\hat X}_{t}\mathbf{\hat W}_{Q}^{m}, \mathbf{K}_{t}^{m} =\mathbf{\hat X}_{t}\mathbf{\hat W}_{K}^{m}, \mathbf{V}_{t}^{m} =\mathbf{\hat X}_{t}\mathbf{\hat W}_{V}^{m}
\end{equation}
As shown in Figure~\ref{fig:SA}, Layer Normalization \cite{ba2016layer} is applied to the input of the two self-attention modules separately. A residual connection \cite{he2016deep} is applied between the input and output of the self-attention module to alleviate the vanishing gradient problem \cite{liang2021attention}.

\subsection{Aggregation module based on graph attention}\label{subsec:ga}
As shown in Figure~\ref{fig:GAT}, we use GAT-aggregation instead of SA-aggregation. GAT-aggregation consists of a cross-frame GAT layer and a cross-channel GAT layer. The detailed calculation process is as follows.
\subsubsection{Cross-frame graph attention layer}
For the cross-frame GAT layer, we first formulate a graph using the frame-level features. Specially, the input of the cross-frame GAT layer is $\mathbf{X} = \left[\mathbf{X}_{0}, \cdots, \mathbf{X}_{C-1}\right]$, where $\mathbf{X}_{c} \in \mathbb{R}^{T \times D}$ and $\mathbf{X}_c = \left[\mathbf{x}_0, \cdots, \mathbf{x}_{T-1}\right]$.
Let $\mathbb{G}_1$ be a complete graph comprising of $T$ nodes. A set of nodes in $\mathbb{G}_1$ is defined as $\mathbf{X}_c$. Each node is represented as a row vector $\mathbf{x}_{t}$, which is a $D$-dimensional feature. For $m$-th attention head, $\mathbf{X}_{c}$ is first projected into a $d_k$-dimensional space as follows:
\begin{equation}
\mathbf{g}_{l}^{m} = \mathbf{X}_{c} \mathbf{W}_{l}^m, \mathbf{g}_{r}^{m} = \mathbf{X}_{c} \mathbf{W}_{r}^m,
\end{equation}
where $\mathbf{g}_{l}^{m},\mathbf{g}_{r}^{m} \in \mathbb{R}^{T \times d_k}$, and $\mathbf{W}_{l}^m, \mathbf{W}_{r}^m \in \mathbb{R}^{D \times d_k}$. Then, we calculate the attention scores by:
\begin{equation}
\begin{gathered}
e_{jk}^m=\beta^\top \operatorname{ LeakyReLU }\left( \operatorname{concat}\left(\mathbf{g}_{li}^m, \mathbf{g}_{rj}^m\right)\right), \\
a_{i j}^m=\operatorname{softmax}\left(e_{i j}^m\right)=\frac{\exp \left(e_{i j}^m\right)}{\sum_{n=1}^{N} \exp \left(e_{i n}^m\right)},
\end{gathered}
\end{equation}
where $a_{ij}^m$ is the attention score between the $i$-th node and the $j$-th node ($\mathbf{g}_{li}^m,\mathbf{g}^m_{rj} \in \mathbb{R}^{d_k}, i \ne j$). $\beta \in \mathbb{R}^{d_k \times 1}$ is a learnable parameter. Then we calculate the output of each head as follows:
\begin{equation}
\mathbf{h}^m_i=\sum_{j=1}^{N} a_{i j}^{m} \mathbf{g}_{rj}^{m}.
\end{equation}
where $\mathbf{h}^m_{i} \in \mathbb{R}^{d_k}$. Finally, we denote $\mathbf{H}^k_c = \left[\mathbf{h}_0^k, \mathbf{h}_1^k, \cdots, \mathbf{h}_{T-1}^k\right] $, and concatenate the outputs of all heads as follows:
\begin{equation}
\mathbf{Y}_{c}=\operatorname{concat} \left[\mathbf{H}_{c}^{1}, \mathbf{H}_{c}^{2}, \ldots, \mathbf{H}_{c}^{M}\right].
\end{equation}
\subsubsection{Cross-channel graph attention layer}
\begin{figure}[t]
	
	\centering
	\includegraphics[width=0.7\linewidth]{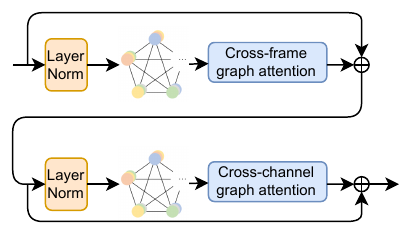}
	\caption{Architecture of the multi-channel aggregation based on graph attention.}
	\label{fig:GAT}
\end{figure}

{We transform $\mathbf{Y} = \left[\mathbf{Y}_{0}, \cdots ,\mathbf{Y}_{C-1}\right]$ to $\mathbf{\hat X} = \left[\mathbf{\hat X}_{0}, \cdots ,\mathbf{\hat X}_{T-1}\right]$ as the input of the cross-channel GAT layer.} Similar to the cross-frame GAT layer, we conduct the following successively steps:
\begin{equation}
\begin{gathered}
\mathbf{g}_{l}^{m} = \mathbf{\hat X}_{t} \mathbf{W}_{l}^m, \mathbf{g}_{r}^{m} = \mathbf{\hat X}_{t} \mathbf{W}_{r}^m, \\
e_{jk}^m=\beta^\top \operatorname{ LeakyReLU }\left( \operatorname{concat}\left(\mathbf{g}_{li}^m, \mathbf{g}_{rj}^m\right)\right), \\
a_{i j}^m=\operatorname{softmax}\left(e_{i j}^m\right)=\frac{\exp \left(e_{i j}^m\right)}{\sum_{n=1}^{N} \exp \left(e_{i n}^m\right)}, \\
\mathbf{h}^m_i=\sum_{j=1}^{N} a_{i j}^{m} \mathbf{g}_{rj}^{m}.
\end{gathered}
\end{equation}
Then, we get the output of the cross-channel GAT layer as follows:
\begin{equation}
\mathbf{Y}_{c}=\operatorname{concat} \left[\mathbf{H}_{t}^{1}, \mathbf{H}_{t}^{2}, \ldots, \mathbf{H}_{t}^{h}\right]
\end{equation}
As shown in Figure \ref{fig:GAT}, similar to SA-aggregation, we also add the Layer Normalization \cite{ba2016layer} before the two graph attention layers separately. The residual connection \cite{he2016deep} is applied between the input and output of the graph attention layers.

\begin{table*}[t]
	
	\begin{center}
		\caption{EER (\%) comparision on the test set of the Libri-adhoc40 corpus.}
\scalebox{0.93}{
		\begin{threeparttable}
			\begin{tabular}{lclllll}
				\toprule
				\multicolumn{1}{c}{\multirow{2}{*}{Method}} & \multicolumn{1}{c}{\multirow{2}{*}{Number of parameters}}& \multicolumn{1}{c}{\multirow{2}{*}{Train set}} & \multicolumn{4}{c}{Test set of Libri-adhoc40} \\
				
				\multicolumn{1}{c}{}         & \multicolumn{1}{c}{}    & \multicolumn{1}{c}{}   & 8 channels     & 16 channels     & 32 channels &40 channels    \\
				\midrule
				\multicolumn{1}{l}{Oracle one-best} & \multicolumn{1}{c}{1.437 M}                                           & \multicolumn{1}{c}{$-$}         & 26.2217       & 21.2862       & 16.1878       & 14.5232              \\
				\multicolumn{1}{l}{Mean pooling} & \multicolumn{1}{c}{1.437 M}                                            & \multicolumn{1}{c}{$-$}        & 20.5158        & 19.7459       & 19.6453       & 19.5376              \\
				\multicolumn{1}{l}{EV \cite{wolf2014channel}}& \multicolumn{1}{c}{1.437 M}                                           & \multicolumn{1}{c}{$-$}       & 24.1581       & 20.5435       & 17.2928       & 15.7186             \\
				\multicolumn{1}{l}{Beamforming \cite{anguera2007acoustic}}&  \multicolumn{1}{c}{1.437 M}                                           & \multicolumn{1}{c}{$-$}      & 19.6441       & 15.8513       & 12.9207       & 12.0551              \\
				\hline
				\multirow{2}{*}{Attentive pooling \cite{cai2021embedding}}& \multirow{2}{*}{1.454 M}                                            & \multicolumn{1}{c}{Simulation}        & 14.4540       & 13.1235       & 12.1465       & 11.9208             \\
				\multicolumn{1}{l}{}& \multicolumn{1}{c}{}                                            & \multicolumn{1}{c}{Real}        & 10.4251       & 9.7030       & 9.2164       & 9.1536             \\
				\hline
				\multirow{2}{*}{UCSA \cite{liang2021attention}}& \multirow{2}{*}{1.503 M}                                            & \multicolumn{1}{c}{Simulation}        & 13.9136       & 12.5733       & 11.2091       & 10.8437             \\
				\multicolumn{1}{l}{}& \multicolumn{1}{c}{}                                            & \multicolumn{1}{c}{Real}        & 10.2522       & 9.4263       & 8.5960       & 8.4308             \\
				\hline
				\multirow{2}{*}{SA-aggregation (proposed)}& \multirow{2}{*}{1.570 M}                                            & \multicolumn{1}{c}{Simulation}        & 13.6264       & 12.6282       & 11.4266       & 11.2482             \\
				\multicolumn{1}{l}{}& \multicolumn{1}{c}{}                                            & \multicolumn{1}{c}{Real}        & \textbf{9.7846}       & 8.5505       & 7.5891      & 7.3640             \\
				\hline
				\multirow{2}{*}{GAT-aggregation (proposed)}& \multirow{2}{*}{1.503 M}                                            & \multicolumn{1}{c}{Simulation}        & 13.2089       & 11.7986       & 10.5637       & 10.2163             \\
				\multicolumn{1}{l}{}& \multicolumn{1}{c}{}                                            & \multicolumn{1}{c}{Real}        & 10.0407       & \textbf{8.3622}       & \textbf{7.1318}       & \textbf{6.8627}             \\
				\bottomrule
			\end{tabular}
		\end{threeparttable}
}
		\label{tab:1}
	\end{center}
\end{table*}

\section{Experiments}
\label{sec:experiments}
\subsection{Dataset}
Our experiments use three data sets, which are the Librispeech corpus \cite{panayotov2015librispeech}, Librispeech simulated with ad-hoc microphone arrays (Libri-adhoc-simu), and Libri-adhoc40 \cite{guan2021libri}. Each node of the ad-hoc microphone arrays of Libri-adhoc-simu and Libri-adhoc40 has only one microphone. Therefore, a channel refers to a node in the remaining of the paper.

The Libri-adhoc-simu corpus is a simulation database of Librispeech. We used 'train-clean-100', 'dev-clean' and 'test-clean' as the training, validation and test sets of the simulated ad-hoc data respectively. The training set contains $251$ speakers. The validation and test sets contain $40$ speakers respectively. We added room impulse response and noise to the clean speech data \cite{liang2021attention}. The size of the simulation room is $[10,10,4]$ meters and the range of the reverberation time is $[0.5-1.2]$ seconds. We randomly placed a speaker and forty microphones in the room. The noise for the training and validation sets was randomly selected from a large-scale noise segments library \cite{tan2021speech}. The noise of the test set comes from the CHiME-3 dataset \cite{barker2015third} and NOISEX-92 \cite{varga1993assessment}. The RIR-Generator\footnote{https://github.com/ehabets/RIR-Generator} and ANF-Generator\footnote{https://github.com/ehabets/ANF-Generator} were used for the data simulation. The Libri-adhoc40 corpus was collected via playing the speech data of Librispeech in a large room, in which forty microphones and a speaker were placed \cite{guan2021libri}.

Considering that a large amount of \textit{data from a massive ad-hoc array}, denoted as ad-hoc data for short, leads to a large memory requirement for the model training, all comparison multi-channel ASV models based on ad-hoc microphone arrays first trained a single-channel ASV with clean speech data, then used the single channel ASV to initialize the multi-channel ASV model, and finally used the ad-hoc data to fine-tune the multi-channel aggregation block. In our experiments, the single channel ASV systems were trained with $960$ hours of the clean data of Librispeech, with another $10$ hours of the clean data was used for development. To study the generalization ability of the proposed systems, we fine-tuned the systems using Libri-adhoc-simu and Libri-adhoc40 respectively.
For each epoch of fine-tuning with a dataset, we randomly selected $20$ channels from the training and validation sets respectively. In the test stage, we only considered the Libri-adhoc40 corpus, and randomly selected $8$, $16$, $32$ and $40$ channels to construct four test scenarios.


\subsection{Experimental setup}

For the proposed SA-aggregation and GAT-aggregation, the network structure of their initial single-channel ASV is the same as in \cite{chung2020defence}, which contains three main components: a front-end residual convolution neural network (ResNet) \cite{he2016deep}, a self-attentive pooling (SAP) \cite{zhu2018self} layer and a fully-connected layer. It was trained for $200$ epochs on the Librispeech corpus. Then, the parameters of the ResNet layer and SAP layer were fixed and sent to the proposed multi-channel ASV. Finally, we fine-tuned the multi-channel aggregation block with Libri-adhoc-simu and Libri-adhoc40 data respectively, where the number of the aggregation block is $1$, and the number of the attention heads is $4$. We used \textit{voxceleb\_trainer}\footnote{https://github.com/clovaai/voxceleb\_trainer} to build our models. The preprocessing of the data and training setting of the proposed models are the same as \cite{liang2021attention}. The following six baselines were used for comparison:
\begin{itemize}
	\item \textbf{Oracle one-best}: We pick the channel that is physically closest to the speaker source as the input of the single-channel ASV model. Note that, for the \textit{oracle one-best} baseline, the distances between the speaker and the microphones are known beforehand.
	\item \textbf{Utterance-level cross-channel self-attention (UCSA) \cite{liang2021attention}}: It adds an utterance-level cross-channel self-attention layer after the pooling layer. Unlike the original UCSA in \cite{liang2021attention}, we removed FFN which leads to better performance than the original UCSA. The number of SA layers and attention heads are $1$ and $4$ respectively.
	\item \textbf{Attentive pooling \cite{cai2021embedding}}: It adds an attentive pooling layer above the pooling layer of the single-channel ASV.
	\item \textbf{Mean pooling}: It sets equal weights to all speaker embeddings from different channels.
	\item \textbf{EV \cite{wolf2014channel}}: It selects the signal of a microphone with the highest envelope variance as the input of the single-channel ASV.
	\item \textbf{Beamforming \cite{anguera2007acoustic}}: It aggregates the multi-channel signals into a single-channel enhanced signal by classic acoustic beamforming. The enhanced speech is taken as the input of the single channel ASV.
\end{itemize}

\subsection{Results}
Tabel~\ref{tab:1} lists the preformance of the comparison methods on the Libri-adhoc40 test set. From the table, we see that, as the number of channels increases, almost all methods achieve better performance. The EV method, which tends to select a channel that is least affected by reverberation, achieves better performance than the \textit{Oracle one-best} method in the $8$-channel and $16$-channel test environments. It indicates that reverberation affects ASV performance. From the results of Beamforming, we find that adding a multichannel speech enhancement front-end is a good way of improving the ASV performance when the number of the channels is large enough. {From the table, we see that the attentive-pooling and UCSA methods trained with the real data outperform the \textit{Oracle one-best}, EV, and Beamforming, which shows that the attention scheme is good at capturing the global information across channels.}

More importantly, from the table, we also see that all of the proposed methods perform well on the Libri-adhoc40 test set. Specially, compared with the attentive-pooling and UCSA, the SA-aggregation fine-tuned with the real data achieves a relative EER reduction of $19.55\%$ over the attentive-pooling, and $12.65\%$ over UCSA in the $40$-channel test scenario. It demonstrates the effectiveness of the frame-level modeling strategy over the utterance-level one. Moreover, GAT-aggregation achieves better performance than SA-aggregation, in the 16-, 32- and 40-channel test scenarios. For example, the GAT-aggregation fine-tuned with the real data achieves a relative EER reduction of $6.02\%$ and $6.81\%$ over SA-aggregation in the 32- and 40-channels test scenarios, respectively. This result shows that the connections between the microphone nodes provide important information for the performance improvement of the multi-channel ASV with a large number of ad-hoc microphone nodes. Although the proposed models fine-tuned with the simulated data are inferior to those fine-tuned with the real data, they still outperform most of the other baseline systems, showing that the proposed methods have a strong generalization ability.

\section{Conclusion}

In this paper, we propose a novel multi-channel ASV framework with ad-hoc microphone arrays. It conducts channel aggregation at the frame-level by adding a multi-channel aggregation module before the pooling layer. We propose two types of multi-channel aggregation modules: SA-aggregation and GAT-aggregation. The aggregation module can be trained in a way that is independent of the number and permutation of the microphones. Experimental results show that the proposed frame-level modules are able to mine more spatial and temporal information than the utterance-level modules. Moreover, GAT-aggregation performs better than SA-aggregation, when the number of nodes of the ad-hoc microphone arrays is large. {Finally, the models fine-tuned with the simulated data have a good generalization ability to the real test scenarios.}

\bibliographystyle{IEEEtran}

\bibliography{myref}


\end{document}